# The Orbit of Transneptunian Binary

# Manwë and Thorondor

# and their Upcoming Mutual Events

W.M. Grundy[a], S.D. Benecchi[b], S.B. Porter[a,c], K.S. Noll[d],

a. Lowell Observatory, 1400 W. Mars Hill Rd., Flagstaff AZ 86001.
b. Planetary Science Institute, 1700 E. Fort Lowell Suite 106, Tucson AZ 85719.
c. Moving to Southwest Research Institute, 1050 Walnut St. #300, Boulder CO 80302.
d. NASA Goddard Space Flight Center, Greenbelt MD 20771.

## Abstract

A new Hubble Space Telescope observation of the 7:4 resonant transneptunian binary system (385446) Manwë has shown that, of two previously reported solutions for the orbit of its satellite Thorondor, the prograde one is correct. The orbit has a period of 110.18 ± 0.02 days, semimajor axis of 6670 ± 40 km, and an eccentricity of 0.563 ± 0.007. It will be viewable edge-on from the inner solar system during 2015-2017, presenting opportunities to observe mutual occultation and eclipse events. However, the number of observable events will be small, owing to the long orbital period and expected small sizes of the bodies relative to their separation. This paper presents predictions for events observable from Earth-based telescopes and discusses the associated uncertainties and challenges.

## Introduction

The Kuiper belt is a region beyond the orbits of the giant planets populated by small icy planetesimals left over from the formation of the solar system. In this zone, transneptunian objects (TNOs) occupy a variety of distinct classes of heliocentric orbits, a dynamical configuration that has been exploited to constrain the early history of the outer solar system. Although the past two decades have seen a spectacularly rapid pace of discovery about the Kuiper belt, the observational challenges of studying the small, distant, and faint TNOs are such that detailed physical and chemical knowledge about individual bodies remains quite sparse. The situation is especially problematic considering that statistical comparisons in these properties between representative samples of different dynamical populations are needed to test hypotheses about the formation and early history of these bodies. The existence of binaries in the various dynamical sub-populations offers a powerful tool for more detailed characterization, beginning with their mutual orbits and dynamical masses. Ideally, mutual events can be observed to determine accurate sizes and thus densities, along with possibilities of determining shapes and mapping surface color and albedo features. For transneptunian binaries with heliocentric orbital periods on the order of multiple centuries, mutual events are rare and valuable occurrences that should be exploited whenever it is possible to do so. So far, they have only been observed in three such systems: Pluto, Haumea, and Sila-Nunam. This paper provides circumstances for observing upcoming mutual events in another transneptunian binary system, (385446) Manwë and its satellite Thorondor.

Manwë was discovered in 2003 by the Deep Ecliptic Survey project (Buie et al. 2003), using the 4 m Blanco telescope at Cerro Tololo, with a confirming follow-up observation nine nights later using the 6.5 m Clay telescope at Las Campanas Observatory. It was given the provisional designation 2003 QW$_{111}$. Its heliocentric orbit, with orbital elements averaged over 10 Myr of $a_\odot = 43.73$ AU, $i_\odot = 1.26°$, and $e_\odot = 0.109$, is in a mean motion resonance with Neptune. For every seven orbits of Neptune around the Sun, Manwë completes four orbits, indicated as the "7:4" resonance (e.g., Gladman et al. 2012). Although TNOs have been discovered in many different mean motion resonances with Neptune, relatively little is known about the physical properties of resonant objects other than those in the comparatively well-studied 3:2 and 2:1 resonances, making this a particularly interesting target for follow-up studies. Lykawka and Mukai (2005) noted that in ($a_\odot$, $e_\odot$, $i_\odot$) space, the 7:4 resonance overlaps the core of the classical Kuiper belt. Their integrations showed that nearby classical TNOs can be influenced by the proximity of the resonance, and that objects can even transition between the two dynamical classes (see also Volk and Malhotra 2011). Exterior to the 7:4 resonance, low inclination classical TNOs have a broader eccentricity distribution, possibly as a result of the outward migration of the 7:4 resonance displacing inner classical TNOs on more eccentric orbits (Morbidelli et al. 2014). CCD photometry of a sample of eleven 7:4 objects revealed most to have very red colors at visible wavelengths (Gulbis et al. 2006; Sheppard 2012), similar to the very red colors prevalent in the dynamically cold core of the classical Kuiper belt (e.g., Gulbis et al. 2006; Peixinho et al. 2008). Besides their colors, the cold classical TNOs are also distinct in having a high rate of binarity (Noll et al. 2008). This characteristic might also be expected to be shared with the 7:4 resonant objects, if objects in that resonance derived from the cold classical region, or if both regions were populated from the same primordial source (e.g., Noll et al. 2012). The Hubble Space Telescope (HST), the leading facility for discovering binary TNOs, has to date observed twelve members of the 7:4 resonant population, but only Manwë was found to be binary when the HST observation revealed a faint companion at a separation of about 0.3 arcsec (Noll et al. 2006). At first glance, finding only one binary out of a dozen seems incompatible with the idea that most 7:4 resonant objects originate from the same source as the cold classical TNOs. Noll et al. (2008) reported a binary rate of 29% among cold classical TNOs. If the binary rate among 7:4 resonant objects was also 29%, the probability of finding one or fewer binaries in a sample of 12 would be about 10%, so the single binary in a sample of 12, while suggestive, does not prove that these objects are different from cold classical TNOs in their binary rate. It would be useful to increase the size of the sample of 7:4 resonant objects observed by HST. In the meantime, the binary nature of the Manwë system opens up a treasure chest of opportunities for more detailed investigations into the physical characteristics of this system, and by extension, the 7:4 resonant population.

## Observations, Photometry, and Orbit Solution

Thorondor, the companion to Manwë, was discovered using HST's Advanced Camera for

Surveys High Resolution Camera (ACS/HRC; Ford et al. 1996) during Cycle 15. That instrument ceased functioning shortly thereafter, so follow-up HST observations to determine the mutual orbit were done using the older WFPC2/PC camera, as part of Cycle 16 program 11178. Grundy et al. (2011) published a pair of Keplerian orbit solutions based on the discovery plus follow-up observations. These two orbit solutions were mirror images of one another through the sky plane, one prograde and one retrograde with respect to Manwë's heliocentric orbit. In Cycle 21 we obtained one more HST orbit to break that mirror ambiguity, as part of program 13404. The observation was executed 2013/11/20 UT, using the UVIS2 camera of the new Wide Field Camera 3 (WFC3; Dressel et al. 2012) that had been installed in place of WFPC2 during the fourth servicing mission to HST. The observing sequence consisted of four dithered integrations through each of the *F438W* and *F606W* filters, broadband filters with nominal central wavelengths of 438 nm and 606 nm, respectively. The four 200 second *F606W* integrations were split into two sets of two, one set just before and one set just after a set of four consecutive, longer, 340 second *F438W* images. This *F606W-F438W-F606W* bookend configuration was designed to limit potential color confusion from lightcurves of the components without adding excessive overhead to the observation sequence.

    Our pipeline for processing the WFC3 images mirrors our processing of ACS/HRC and WFPC2/PC data described in previous publications, so we refer interested readers to those papers for more details (e.g., Benecchi et al. 2009; Grundy et al. 2009, 2011, 2012). Briefly, for each separate frame, we fitted a pair of Tiny Tim model point spread functions (Krist and Hook 2004; Krist et al. 2011) to the two sources, then used the scatter in the modeled positions and fluxes between the separate frames to estimate the uncertainties in our measurements of those parameters. Table 1 shows the measured relative astrometry for our new WFC3 observation along with earlier observations and Table 2 shows photometric brightnesses of the components

**Table 1**
**Observations of Astrometry for Thorondor relative to Manwë**

| UT date and time | Instrument/camera | $r$[a] | $\Delta$[a] | $g$[a] | $\Delta x$ | $\Delta y$ |
|---|---|---|---|---|---|---|
| | | (AU) | | (deg.) | (arcsec)[b] | |
| 2006/07/25  9$^h$.1349 | ACS/HRC | 44.743 | 43.949 | 0.82 | +0.2948(10) | –0.1348(14) |
| 2007/07/25  3$^h$.6392 | WFPC2/PC | 44.643 | 43.868 | 0.85 | +0.1556(21) | –0.0979(10) |
| 2007/08/26 13$^h$.7017 | WFPC2/PC | 44.634 | 43.637 | 0.21 | +0.028(38) | –0.023(44) |
| 2008/08/04 19$^h$.2138 | WFPC2/PC | 44.539 | 43.666 | 0.67 | +0.1606(25) | –0.0501(26) |
| 2008/08/20 15$^h$.5804 | WFPC2/PC | 44.535 | 43.559 | 0.35 | +0.2619(16) | –0.1093(13) |
| 2008/09/07 14$^h$.0936 | WFPC2/PC | 44.530 | 43.524 | 0.08 | +0.3000(40) | –0.1345(40) |
| 2008/10/26 19$^h$.5369 | WFPC2/PC | 44.516 | 43.893 | 1.00 | –0.0756(64) | +0.0208(38) |
| 2013/11/20 16$^h$.5083 | WFC3/UVIS2 | 43.981 | 43.640 | 1.21 | +0.1964(27) | –0.1057(39) |

Table notes:

[a.] The distance from the Sun to the target is $r$ and from the observer to the target is $\Delta$. The phase angle $g$ is the angular separation between the observer and Sun as seen from the target.

[b.] Relative right ascension $\Delta x$ and relative declination $\Delta y$ are computed as $\Delta x = (\alpha_2 - \alpha_1)\cos(\delta_1)$ and $\Delta y = \delta_2 - \delta_1$, where $\alpha$ is right ascension, $\delta$ is declination, and subscripts 1 and 2 refer to Manwë and Thorondor, respectively. Estimated 1-$\sigma$ uncertainties in the final digits are indicated in parentheses. Uncertainties are estimated from the scatter between fits to individual frames, except for 2008/09/07, when only a single usable frame was obtained and 4 mas uncertainties were assumed.

from the available HST observations.

<table>
<caption>Table 2<br>Separate Photometry for Manwë and Thorondor</caption>

| UT Date | Manwë | | | | Thorondor | | | | | $\Delta_{mag}$ |
|---|---|---|---|---|---|---|---|---|---|---|
| | Clear | B | V | $H_V$ | I | Clear | B | V | $H_V$ | I | |
| 2006/07/25 | 23.491(13) | - | - | | - | 25.57(22) | - | - | | - | 2.08(22) |
| 2007/07/25 | - | - | 24.388(85) | 7.789(85) | - | - | - | 24.970(49) | 8.371(49) | - | 0.58(10) |
| 2007/08/26 | - | - | 24.098(35) | 7.592(35) | - | - | - | 26.18(15) | 9.67(15) | - | 2.09(15) |
| 2008/08/04 | - | - | 23.849(43) | 7.285(43) | 22.686(29) | - | - | 25.26(12) | 8.70(12) | 23.950(73) | 1.305(67) |
| 2008/08/20 | - | - | 23.870(48) | 7.352(48) | 22.598(42) | - | - | 25.22(10) | 8.70(10) | 24.098(84) | 1.438(73) |
| 2008/09/07 | - | - | 24.22(10) | 7.75(10) | - | - | - | 24.80(20) | 8.33(20) | - | 0.58(22) |
| 2008/10/26 | - | - | 24.09(13) | 7.48(13) | - | - | - | 24.97(11) | 8.36(11) | - | 0.88(17) |
| 2013/11/20 | - | 24.911(57) | 23.855(54) | 7.265(54) | - | - | 25.500(47) | 24.575(36) | 7.985(36) | - | 0.664(66) |

</table>

Table note
Photometric uncertainties were estimated from the scatter between multiple frames, except for 2008/09/07 when only a single frame was available. Photometry was converted from HST filters *F438W*, *F606W*, and *F814W* into Johnson *B*, *V*, and *I* magnitudes using synphot as described in detail by Benecchi et al. (2009). Magnitude differences $\Delta_{mag}$ between Manwë and Thorondor are computed from all filters used on each date. Absolute magnitudes $H_V$ are derived from the *V* photometry by assuming *G* =0.15 in the Bowell et al. (1989) photometric system.

Even discounting the 2007/08/26 observation when the two objects were highly blended, differences between Manwë and Thorondor magnitudes in Table 2 are quite diverse, with $\Delta_{mag}$ ranging from 2.1 ± 0.2 on 2006/07/25 to 0.6 ± 0.1 on 2007/07/25 (and also on 2008/09/07). Evidently this system exhibits considerable photometric variability over time. If we attribute all such variability to irregular shapes rather than potentially wavelength-dependent albedo/color patterns, we can combine $\Delta_{mag}$ values from dissimilar filters. In addition to the *F435W* and *F606W* filters, the system was resolved through other HST filters *F814W* (nominal central wavelength 814 nm) and *CLEAR* (unfiltered). Weighting each epoch equally, we obtain an average $<\Delta_{mag}>$ = 1.2. Although the photometric variability is inconsistent with such a scenario, if both bodies were spheres with a common albedo, this $\Delta_{mag}$ would correspond to a radius of Thorondor about 58% that of Manwë and Thorondor would comprise about 16% of the total system volume and mass, if both objects had the same bulk density.

On three of our observation dates, the system was imaged through a sequence of two different filters, enabling us to compare the colors of Manwë and Thorondor. From the photometry in Table 2, we find *V−I* colors of 1.163 ± 0.057 and 1.31 ± 0.14 mags for Manwë and Thorondor on 2008/08/04, and 1.272 ± 0.064 and 1.12 ± 0.13 on 2008/08/20. The *B−V* colors on 2013/11/20 are 1.056 ± 0.079 and 0.925 ± 0.059 mags for Manwë and Thorondor. Subtracting Thorondor's colors from Manwë's on each of the three dates, we get color differences of −0.15 ± 0.15, +0.15 ± 0.14, and +0.131 ± 0.099 mags, all statistically indistinguishable from no color difference between the two, consistent with the Benecchi et al. (2009) finding that the primary and secondary bodies of transneptunian binaries tend to share a common color. Merging the fluxes of Manwë and Thorondor to get colors for the full system, we obtain *V−I* colors of 1.196 ± 0.051 mags on 2008/08/04 and 1.240 ± 0.057 mags on 2008/08/20 and a *B−V* color of 1.010 ± 0.055 mags on 2013/11/20. These are comparable to the very red colors found for 7:4 resonant objects by Sheppard (2012). For Manwë and Thorondor, that paper reported *V−R* = 0.61 ± 0.06 mags, *B−R* = 1.68 ± 0.07 mags, and *R−I* = 0.61 ± 0.04 mags, from which we can compute $V-I$ = 1.22 ± 0.07 mags and $B-V$ = 1.07 ± 0.09 mags, statistically indistinguishable from our colors.

We can compensate for effects of time-variable geometry between the photometric measurements in Table 2 by assuming generic asteroid-like photometric behavior with $G$ = 0.15 in the $H$ and $G$ system of Bowell et al. (1989) to convert all of our $V$ magnitudes to absolute magnitudes $H_V$, as listed in Table 2. However, the suitability of this $G$ value for this particular system is unknown. TNOs show considerable diversity in their photometric behaviors and smaller, more distant objects such as this one tend to be under-represented in studies of TNO phase functions (e.g., Rabinowitz et al. 2007; Belskaya et al. 2008; Stansberry et al. 2008). Weighting each of our seven epochs equally, along with an additional $V$ magnitude computed from the $V-R$ and $R$ photometry of Sheppard (2012), we estimate the time-averaged absolute magnitude of the combined system as $H_V$ = 7.15, although our sparse and non-random temporal sampling leaves a lot to be desired. The minimum, median, and maximum $H_V$ values were 6.81, 7.23, and 7.44, respectively.

From the separate $H_V$ photometry in Table 2, both Manwë and Thorondor appear to show substantial photometric variability that is not obviously correlated between the two objects, implying they are not tidally locked into a shared rotation state, at least if the variability is attributed to shape effects, rather than albedo markings. Discounting the highly blended 2007/08/26 observation, the photometric observations require peak-to-peak lightcurve amplitudes of at least 0.5 mags for Manwë and 0.7 mags for Thorondor. To create such high amplitude lightcurve variations, their shapes would have to be moderately elongated, or more speculatively, the individual component bodies could themselves be unresolved near-contact or contact binaries. Lightcurve amplitudes of 0.5 and 0.7 mags corresponding to changes in projected area of rotating prolate ellipsoids having long axes greater than their short axes by factors of at least 1.6 and 1.9, respectively. We examined the individual frames for evidence of shorter-term photometric variability that could potentially be indicative of rapid rotation, but within each visit, the frame-to-frame variation was consistent with noise.

Our orbit determination procedure was described in prior publications (e.g., Grundy et al. 2009, 2011, 2012). To find the set of Keplerian orbital elements that minimizes $\chi^2$ for the astrometry in Table 1, we used the downhill simplex algorithm Amoeba (Nelder and Mead 1965; Press et al. 1992). Including the new 2013 data point, the best fit retrograde solution has $\chi^2$ = 30, so it can be excluded at greater than 3-σ confidence, assuming each of the eight observations provides two independent constraints (the relative right ascension $\Delta x$ and declination $\Delta y$ from Table 1) and that the observational errors obey a Gaussian distribution that is accurately described by our tabulated 1-σ error bars on $\Delta x$ and $\Delta y$. The prograde solution has $\chi^2$ = 7 corresponding to a reduced $\chi_v^2$ = 0.8, suggesting that, if anything, we may have slightly over-estimated our astrometric uncertainties. To assess the uncertainties associated with our fitted orbital parameters, we generated a new orbit by adding Gaussian random noise to each observed data point consistent with its error bar, and redid the fitting procedure to find the lowest $\chi^2$ orbit solution for that particular realization of the observational data plus noise. This procedure was repeated 1000 times to accumulate a collection of 1000 randomized orbits consistent with the observational data. This Monte Carlo cloud of orbits was used to determine error bars on the fitted parameters as well as on derived parameters such as the system mass. The resulting orbit, derived parameters, and uncertainties appear in Table 3 and the data, prograde and retrograde orbit solution, and residuals are shown in Fig. 1.

Dynamically, this system is near the transition between where solar and mutual tidal perturbations are most important. Solar tides can cause the orbit to undergo Kozai Cycles, which would periodically increase the orbit's eccentricity (e.g., Kozai 1962; Perets and Naoz 2009; Naoz et al. 2010). At higher eccentricities (and thus closer periapse passages), internal tides on the bodies become more important, allowing for tidal dissipation of energy and transfer of angular momentum between the orbit and the spins of the objects. In addition, close periapse passages allow perturbations from the shape of the objects to become prominent. This system is just close enough that the oblateness of the primary object (its $J_2$ gravity term) could cancel out the solar Kozai cycles (e.g., Nicholson et al. 2008), thus preventing the eccentricity from increasing enough to allow significant tidal decay. Porter and Grundy (2012) showed that a system balanced like this could be stable for the lifetime of the solar system. The magnitude of the oblateness perturbation is a degenerate function of oblateness of the primary, the direction of the spin pole of the primary, and the internal tidal physics of both objects.

**Table 3**
**Mutual Orbit Solution and 1-$\sigma$ Uncertainties**

| Parameter | | Value |
|---|---|---|
| **Fitted elements**[a] | | |
| Period (days) | $P$ | 110.176 ± 0.018 |
| Semimajor axis (km) | $a$ | 6674 ± 41 |
| Eccentricity | $e$ | 0.5632 ± 0.0070 |
| Inclination[b] (deg) | $i$ | 25.58 ± 0.23 |
| Mean longitude[b] at epoch[c] (deg) | $\epsilon$ | 126.51 ± 0.49 |
| Longitude of asc. node[b] (deg) | $\Omega$ | 163.56 ± 0.78 |
| Longitude of periapsis[b] (deg) | $\varpi$ | 250.8 ± 1.9 |
| **Derived parameters** | | |
| Standard gravitational parameter $GM_{sys}$ (km$^3$ day$^{-2}$) | $\mu$ | 0.1295 ± 0.0024 |
| System mass ($10^{18}$ kg) | $M_{sys}$ | 1.941 ± 0.036 |
| Orbit pole right ascension[b] (deg) | $\alpha_{pole}$ | 73.56 ± 0.79 |
| Orbit pole declination[b] (deg) | $\delta_{pole}$ | 64.42 ± 0.24 |
| Orbit pole ecliptic longitude[d] (deg) | $\lambda_{pole}$ | 80.61 ± 0.46 |
| Orbit pole ecliptic latitude[d] (deg) | $\beta_{pole}$ | 41.52 ± 0.24 |

Table notes:
[a.] Elements are for Thorondor relative to Manwë. Excluding the 2007/08/26 observation (effectively a non-detection of Thorondor resulting in a residual of 41 mas), the average sky plane residual is 2.8 mas and the maximum is 6.6 mas; $\chi^2$ is 7.08, based on observations at 8 epochs.
[b.] Referenced to J2000 equatorial frame.
[c.] The epoch is Julian date 2454400.0 (2007 October 26 12:00 UT).
[d.] Referenced to J2000 ecliptic frame.

We simulated the system using the Porter and Grundy (2012) dynamical model with reasonable assumptions for the tidal parameters and oblateness and a variety of spin rates and poles. The simulations did not evolve towards a synchronous state, and were not stable when forced to a synchronous state. Instead, the stable configurations all had spin poles for Manwë which were inclined sufficiently for the oblateness perturbation to be significant. This suggests that Manwë's real rotational pole is probably inclined to the mutual orbit pole by at least 20°, and its rotation rate is unlikely to be synchronized to the orbital period. Because Thorondor is much smaller than Manwë, it has less effect on the mutual orbit, and from the photometric variability discussed earlier, it probably has an even less spherical shape than Manwë. The Porter and Grundy (2012) model cannot simulate an elongated object. However, at an eccentricity of 0.56, Equation 4 in Wisdom et al. (1984) shows that for a triaxial ellipsoid secondary with principal moments of inertia $A \leq B \leq C$, avoiding instability from overlapping 1:1 and 3:2 spin-orbit resonances would require $(B-A)/C < 0.014$, inconsistent with an elongated shape. Thus,

Thorondor is likely to be rotating chaotically, unless it is spinning extremely rapidly.

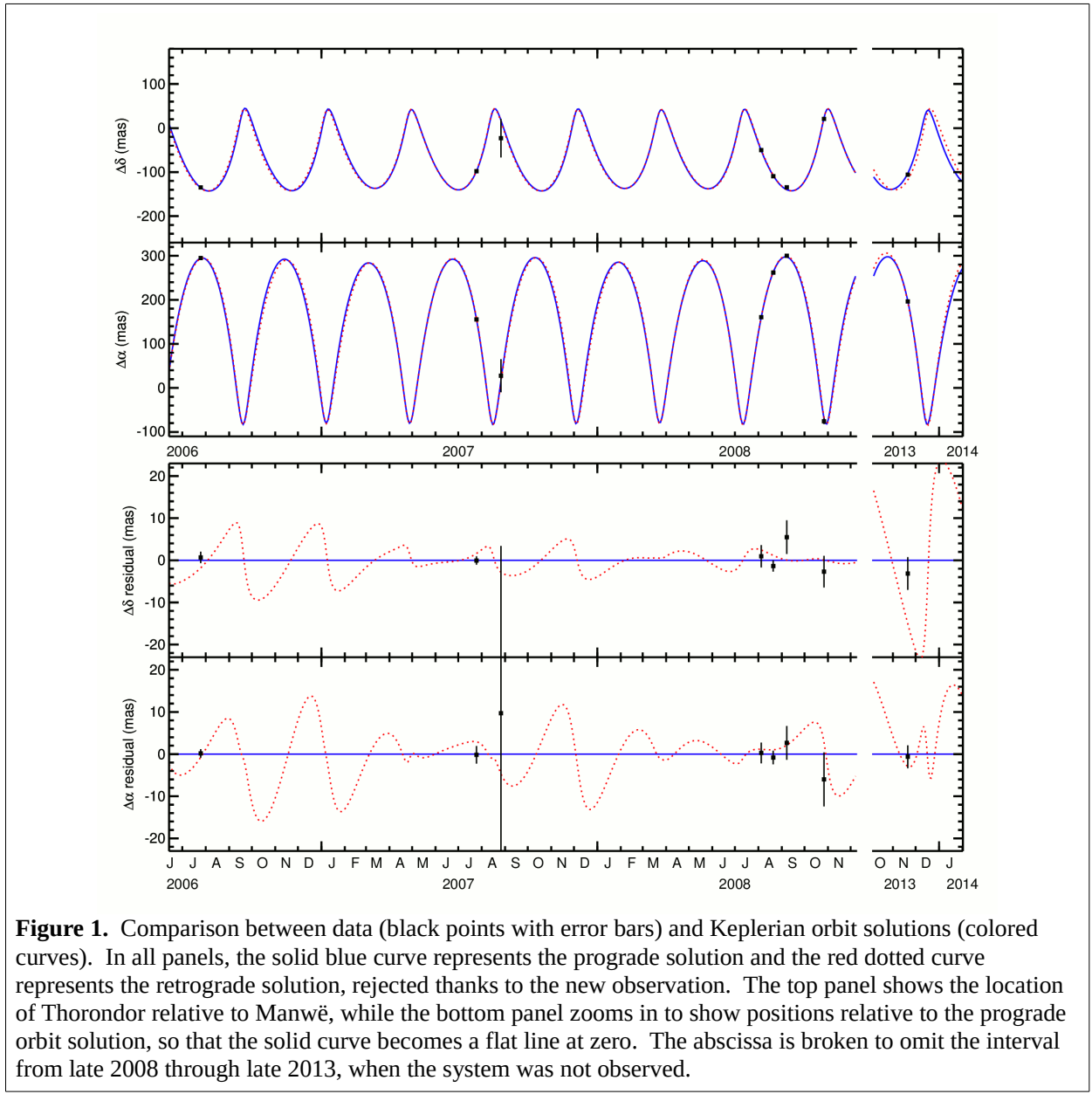

**Figure 1.** Comparison between data (black points with error bars) and Keplerian orbit solutions (colored curves). In all panels, the solid blue curve represents the prograde solution and the red dotted curve represents the retrograde solution, rejected thanks to the new observation. The top panel shows the location of Thorondor relative to Manwë, while the bottom panel zooms in to show positions relative to the prograde orbit solution, so that the solid curve becomes a flat line at zero. The abscissa is broken to omit the interval from late 2008 through late 2013, when the system was not observed.

## Mutual Event Predictions

To forecast mutual event circumstances, we first need to assume sizes, shapes, and center-to-limb photometric behaviors for Manwë and Thorondor. For simplicity, we assume Lambertian photometric behavior and spherical shapes. Although spherical shapes are inconsistent with the apparent lightcurve variability discussed in the previous section, without actually knowing the spin state of either body, it would be premature to employ non-spherical shape models. No size constraints for the bodies have been published to date from the usual methods of thermal radiometry or stellar occultations. Grundy et al. (2011) estimated plausible radius ranges for binaries based on a plausible bulk density range (taken to be between 0.5 and 2.0 g cm$^{-3}$), the system masses, and assumptions of spherical shapes and equal albedos for Manwë and Thorondor. With our new observation, the system mass is revised slightly to $M_{sys}$ = (1.941 ± 0.036) × $10^{18}$ kg and as discussed in the previous section, we have also revised the average $\Delta_{mag}$ to 1.2. Using equations 3 from Grundy et al. (2011)[1], we can then update the plausible radius range to be between 58 and 92 km for Manwë and between 33 and 53 km for Thorondor. This range of plausible sizes can be combined with our mean $H_V$ = 7.15 to obtain a range of plausible geometric albedos between 0.06 and 0.14, unremarkable for a small transneptunian object (e.g., Stansberry et al. 2008; Santos-Sanz et al. 2012; Vilenius et al. 2012), although it is perhaps noteworthy as the only estimated albedo for an object in the 7:4 mean motion resonance. Between the adopted large and small size limits, we also adopt a nominal size case of 80 and 46 km radii for Manwë and Thorondor, respectively, corresponding to a bulk density of 0.75 g cm$^{-3}$. This density value was chosen because the handful of other small transneptunian objects with reasonably well constrained densities fall in the range between 0.5 and 1 g cm$^{-3}$ (e.g., Stansberry et al. 2012; Brown 2013).

### *Best-fit solution*

With these assumptions and the best-fit mutual orbit solution from the previous section, we can project the system as it would appear to an Earth-based observer at a given time. To do this, we combine the geometry within the Manwë – Thorondor system according to our mutual orbit with the position of the system's barycenter relative to Earth and Sun according to JPL's Navigation and Ancillary Information Facility (NAIF) SPICE ephemeris utilities. Example snapshots appear in Figure 2. This plot shows the relative positions of Manwë and Thorondor as seen from Earth on dates of seven potential mutual events when the system is at a relatively large solar elongation. Arrows indicate the sky-plane motion of Thorondor relative to Manwë over the course of 16 hours. Longer arrows indicate more rapid relative motion during inferior events, when Thorondor is in the foreground. The difference in apparent sky-plane relative rates between superior and inferior events is due to the shape and orientation of the eccentric mutual orbit. When the foreground object clips the background object, an occultation event occurs. The shadow of the foreground object at the distance of the background one is indicated by a hatched area. If this shadow clips the background object, an eclipse event occurs.

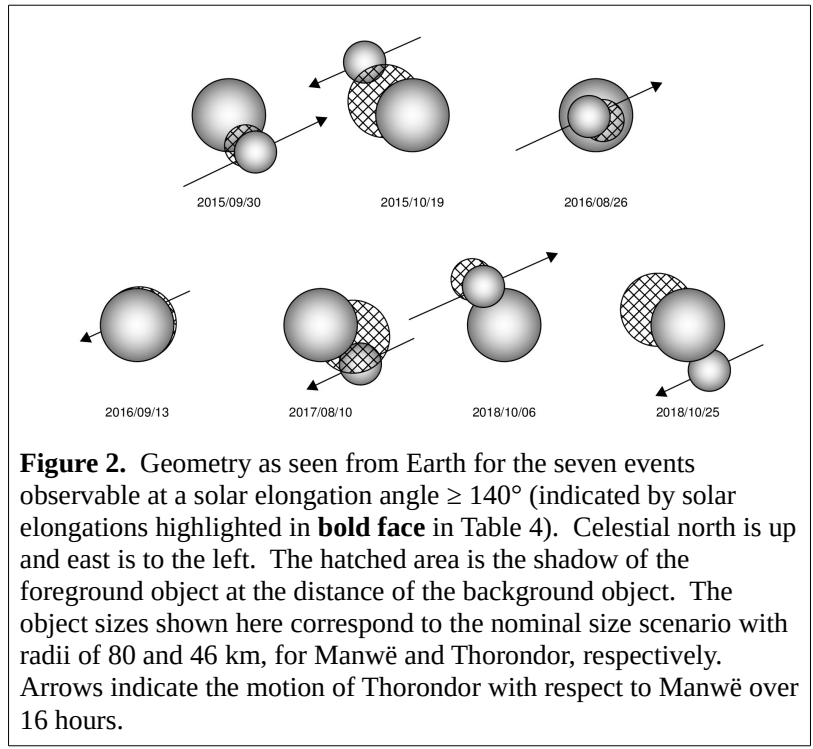

**Figure 2.** Geometry as seen from Earth for the seven events observable at a solar elongation angle ≥ 140° (indicated by solar elongations highlighted in **bold face** in Table 4). Celestial north is up and east is to the left. The hatched area is the shadow of the foreground object at the distance of the background object. The object sizes shown here correspond to the nominal size scenario with radii of 80 and 46 km, for Manwë and Thorondor, respectively. Arrows indicate the motion of Thorondor with respect to Manwë over 16 hours.

---

1  Note that the radii in Table 12 of Grundy et al. (2011) are inconsistent with their equation. The numbers in the table are in error. The equation is correct.

Often both types of events occur together. The location of the shadow depends only on the mutual orbit and the motion of the system around the Sun, but the apparent position of the foreground object relative to its shadow and the background object depends on where Earth is in its heliocentric orbit. Prior to opposition in mid-September, objects appear east of their shadows, whereas after opposition they appear west of their shadows. Near opposition, the shadow is mostly hidden behind the body casting it. Event predictions are also available at our web site (http://www2.lowell.edu/users/grundy/tnbs/385446_2003_ QW111_Manwe-Thorondor_mutual_events.html), and will be updated there as additional information becomes available.

The larger the assumed sizes, the more sky area is swept out by the two bodies, resulting in more events and longer events over a longer mutual event season. For the large size scenario (92 and 53 km for Manwë and Thorondor radii, respectively), all event dates when solar elongation is $\geq 70°$ are listed in Table 4 (next page). We also list information for the nominal and small size scenarios, although they do not produce events on all of the dates. For example, consider the event on 2014/08/04 UT. For the small and nominal body size scenarios, no event occurs. However, an extremely shallow occultation event is observable on this date for the large size scenario.

Event durations depend on both the relative rate of motion (indicated by the arrows in Fig. 2) and on the sizes and geometric configurations of the bodies and their shadows. Durations in hours between first and last contact are listed in Table 4. Some events are quite long, up to 22 hours, especially for the larger size scenarios and for the slower-moving superior events. Clearly, individual Earth-based observing sites cannot expect to monitor the entirety of these events, since the system can only be observed for a limited number of nighttime hours from a single site. Many of the events would require coverage from multiple telescopes located at different longitudes, especially if a reasonable sample of pre- and post-event background signal is desired. Airborne and space-based telescopes have greater flexibility in being able to observe at specific times, so they could make a valuable contribution to this effort.

### *Uncertainties*

So far, we have been considering only the best-fit orbit solution. As described in the previous section, the orbital elements have associated uncertainties. Monte Carlo techniques were used to assess those uncertainties by means of a cloud of 1000 orbit solutions consistent with the observations and their astrometric uncertainties. That same cloud of orbits can be used to investigate the effects of orbit uncertainties on the mutual events. For each date in Table 4, we computed first and last contact times along with the mid-time for each of the 1000 orbits for which an event occurs. The 1-σ scatter of the mid-times is listed as an uncertainty on the event mid-time in hours in the first column of the table. These timing uncertainties gradually grow over time, as the uncertainty in orbital longitude grows, but even at the beginning of the mutual event season they are ten or more hours. These large timing uncertainties compound the already challenging problem of temporal coverage from Earth-based telescopes. Fortunately, the timing uncertainties can be collapsed by obtaining new data closer to the time of the events. The additional data could be relative astrometry, just like the data used to compute the mutual orbit. It could also be a successful observation of an event itself, as was recently used to collapse the timing uncertainties on mutual events in the Sila-Nunam system (Benecchi et al. 2014).

**Table 4**
**Mutual Event Predictions for Manwë and Thorondor**

| Event mid-time[a] | Elongation[b] Solar | Elongation[b] Lunar | Event type[c] | Event probabilities, components, and durations[d] Large | Nominal | Small |
|---|---|---|---|---|---|---|
| 2014/07/16 20 ± 10 | 123 | 7 | Inf. | 1.0, occ., 10 ± 2 | 0.9, occ., 7 ± 2 | - |
| 2014/08/04  6 ± 12 | 141 | 127 | Sup. | 0.5, occ., 2 ± 3 | - | - |
| 2014/11/03 23 ± 10 | 128 | 14 | Inf. | 0.7, ecl., 5 ± 2 | - | - |
| 2015/06/12 11 ± 12 | 89 | 36 | Inf. | 1.0, ecl., occ., 14 ± 1 | 1.0, ecl., occ., 12 ± 1 | 1.0, occ., 8 ± 1 |
| 2015/06/30 24 ± 13 | 106 | 89 | Sup. | 1.0, ecl., occ., 18 ± 2 | 1.0, ecl., occ., 15 ± 2 | 1.0, occ., 10 ± 2 |
| 2015/09/30 14 ± 12 | **164** | 51 | Inf. | 1.0, ecl., occ., 11 ± 1 | 1.0, ecl., occ., 9 ± 2 | 0.7, ecl., 5 ± 2 |
| 2015/10/19  1 ± 13 | **146** | 78 | Sup. | 1.0, ecl., 13 ± 3 | 0.8, ecl., 10 ± 3 | - |
| 2016/05/26 14 ± 14 | 72 | 54 | Sup. | 1.0, ecl., occ., 22 ± 1 | 1.0, ecl., occ., 19 ± 1 | 1.0, ecl., occ., 14 ± 1 |
| 2016/08/26  5 ± 13 | **160** | 84 | Inf. | 1.0, ecl., occ., 14 ± 1 | 1.0, ecl., occ., 12 ± 1 | 1.0, ecl., occ., 9 ± 1 |
| 2016/09/13 18 ± 14 | **177** | 42 | Sup. | 1.0, ecl., occ., 18 ± 1 | 1.0, ecl., occ., 16 ± 1 | 1.0, ecl., occ., 11 ± 1 |
| 2016/12/14  9 ± 13 | 89 | 96 | Inf. | 1.0, ecl., occ., 15 ± 1 | 1.0, ecl., occ., 13 ± 1 | 1.0, ecl., occ., 10 ± 1 |
| 2017/01/01 22 ± 15 | 70 | 29 | Sup. | 1.0, ecl., occ., 21 ± 1 | 1.0, ecl., occ., 18 ± 2 | 1.0, ecl., occ., 13 ± 2 |
| 2017/07/22 20 ± 14 | 125 | 117 | Inf. | 1.0, ecl., occ., 14 ± 1 | 1.0, ecl., occ., 11 ± 1 | 1.0, ecl., 7 ± 2 |
| 2017/08/10 11 ± 15 | **143** | 5 | Sup. | 1.0, ecl., occ., 17 ± 3 | 1.0, ecl., occ., 14 ± 3 | 0.8, ecl., 8 ± 3 |
| 2017/11/09 23 ± 15 | 126 | 133 | Inf. | 1.0, ecl., occ., 15 ± 1 | 1.0, ecl., occ., 13 ± 1 | 1.0, ecl., occ., 9 ± 1 |
| 2017/11/28 13 ± 16 | 107 | 5 | Sup. | 1.0, ecl., occ., 21 ± 2 | 1.0, ecl., occ., 18 ± 2 | 1.0, ecl., occ., 12 ± 2 |
| 2018/06/18 11 ± 15 | 91 | 155 | Inf. | 1.0, ecl., 9 ± 2 | 0.9, ecl., 7 ± 2 | - |
| 2018/07/07  3 ± 16 | 109 | 29 | Sup. | 0.5, ecl., 4 ± 4 | - | - |
| 2018/10/06 15 ± 16 | **162** | 165 | Inf. | 1.0, ecl., occ., 10 ± 2 | 0.9, ecl., occ., 7 ± 2 | - |
| 2018/10/25  6 ± 16 | **143** | 44 | Sup. | 0.9, occ., 12 ± 4 | 0.7, occ., 8 ± 3 | - |
| 2019/12/21  9 ± 17 | 87 | 148 | Inf. | 0.7, occ., 5 ± 2 | - | - |

Table Notes
[a.] UT date and hour midway between first and last contact for the large size scenario, with 1-σ timing uncertainties arising from uncertainties in the orbital elements.
[b.] Solar elongation angle is the angle between the Sun and object as seen from Earth in degrees. Events at elongations less than 70° were excluded since they would be especially difficult to observe from Earth. Elongations ≥ 140°, highlighted with **bold face**, are better candidates for observation from terrestrial telescopes, although some are compromised by proximity of the moon. These seven events are the ones shown in Figs. 2 and 3.
[c.] Event types are indicated by "Sup." for superior events in which Manwë is in front and "Inf." for inferior events in which Thorondor is in front, as seen from Earth.
[d.] Probability of an event is based only on uncertainties in the mutual orbital elements. Events are only shown where probability is at least 50%. Eclipse events in which one body casts a shadow on the other are indicated by "ecl.". Occultation events where one body obstructs the view of the other are indicated by "occ.". Many events involve both eclipse and occultation components. Event durations between first and last contact are given in hours with 1-σ uncertainties due to the orbital elements but not other factors such as size, shape, rotation state, etc.

In addition to the timing uncertainties, uncertainties in the orbital elements can also introduce uncertainty about whether or not an event will even occur. By considering the fraction of the orbit cloud that produces a particular event, we can estimate probabilities for each event. These probabilities are listed in Table 4 to the nearest tenth. As the assumed object sizes are reduced from large to nominal to small, the probabilities decline. Where less than 50% of the orbits produce an event, we replace the event information with a dash, indicating a probable non-event.

Simulated lightcurves are shown in Fig. 3, illustrating effects of the three different size scenarios on event depth, duration, and lightcurve shape. These models give an idea of the sort of signal precision that would be required to determine the actual sizes of the bodies.

Size is not the only parameter that affects these lightcurves. Center-to-limb photometric behavior and orbital elements also influence the depths and durations of events, so in interpreting observational data, it will be necessary to simultaneously solve for object sizes, photometric behavior, and more precise orbital elements. The shapes and rotation states of the bodies could add additional uncertainties to event durations, depending on the orientations of the potentially elongated (or even multiple) objects at the time of each event. This problem is not expected to effect interpretation of Sila-Nunam mutual events, since there are dynamical arguments and evidence from lightcurves that both bodies in that system rotate synchronously with their mutual orbit (e.g., Grundy et al. 2012; Rabinowitz et al. 2014). But for Manwë and Thorondor, elongated shapes could make events longer, if the long axis of either body happens to lie along the sky plane direction of motion during an event. If a long axis is perpendicular to this motion, but still in the sky plane, events will be shorter, but more area on the sky is swept out, so there is a possibility of events occurring on dates not listed in Table 3. Event lightcurves will be more complicated if either body spins fast enough to exhibit appreciable lightcurve variation of its own during the course of a mutual event. Considering all these factors, mutual event observations would be very difficult to interpret without knowledge of the objects' rotation states from additional spatially resolved photometric observations or else unresolved lightcurve studies with sufficient duration and signal precision to enable solving for the photometric contributions of the two bodies. If Thorondor is rotating chaotically, as seems probable, that will compound the difficulty of accounting for its non-spherical shape and orientation still further.

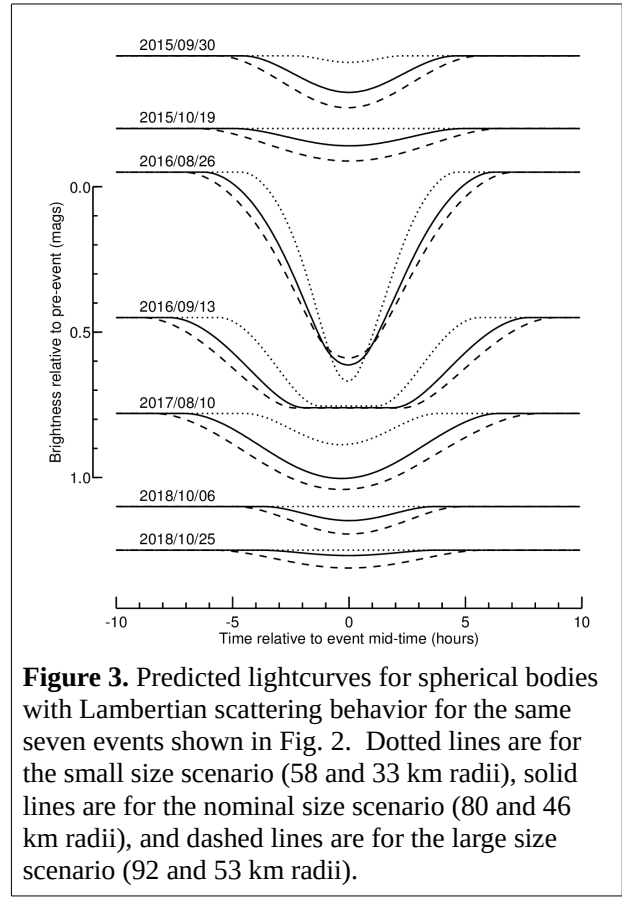

**Figure 3.** Predicted lightcurves for spherical bodies with Lambertian scattering behavior for the same seven events shown in Fig. 2. Dotted lines are for the small size scenario (58 and 33 km radii), solid lines are for the nominal size scenario (80 and 46 km radii), and dashed lines are for the large size scenario (92 and 53 km radii).

**Table 5**
**Additional Events[a] Enabled by Larger Sizes and/or Smaller Solar Elongations**

| Event mid-time | Elongation | | Event type |
|---|---|---|---|
| | Solar | Lunar | |
| 2012/06/05 10:00 | 86 | 80 | Inf. |
| 2012/06/23 16:00 | 104 | 150 | Sup. |
| 2012/09/23 12:00 | 167 | 68 | Inf. |
| 2013/01/11 17:00 | 56 | 57 | Inf. |
| 2013/05/02 00:00 | 52 | 45 | Inf. |
| 2013/05/20 09:00 | 70 | 171 | Sup. |
| 2013/08/20 04:00 | 158 | 35 | Inf. |
| 2013/09/07 10:00 | 175 | 160 | Sup. |
| 2013/12/08 08:00 | 92 | 20 | Inf. |
| 2013/12/26 15:00 | 73 | 150 | Sup. |
| 2014/03/28 14:00 | 20 | 13 | Inf. |
| 2014/04/16 00:00 | 35 | 136 | Sup. |
| 2014/07/16 20:00 | 123 | 7 | Inf. |
| 2014/08/04 05:00 | 141 | 127 | Sup. |
| 2014/11/22 08:00 | 109 | 112 | Inf. |
| 2015/02/22 04:00 | 18 | 27 | Inf. |
| 2015/03/12 14:00 | 3 | 103 | Sup. |
| 2016/01/18 18:00 | 53 | 60 | Inf. |
| 2016/02/06 06:00 | 35 | 65 | Sup. |
| 2016/05/08 01:00 | 54 | 72 | Inf. |
| 2017/04/03 15:00 | 20 | 108 | Inf. |
| 2017/04/22 06:00 | 38 | 19 | Sup. |
| 2018/02/28 05:00 | 15 | 141 | Inf. |
| 2018/03/18 22:00 | 4 | 20 | Sup. |
| 2019/01/24 19:00 | 50 | 178 | Inf. |
| 2019/02/12 12:00 | 32 | 53 | Sup. |
| 2019/05/15 02:00 | 57 | 172 | Inf. |
| 2019/06/02 20:00 | 74 | 67 | Sup. |
| 2019/09/02 07:00 | 162 | 157 | Inf. |
| 2019/09/21 02:00 | 177 | 79 | Sup. |
| 2020/01/09 03:00 | 68 | 91 | Sup. |
| 2020/04/09 17:00 | 23 | 135 | Inf. |
| 2020/07/28 22:00 | 127 | 124 | Inf. |
| 2020/11/16 00:00 | 123 | 112 | Inf. |
| 2020/12/04 19:00 | 104 | 127 | Sup. |
| 2021/03/06 07:00 | 13 | 99 | Inf. |
| 2021/10/12 17:00 | 159 | 75 | Inf. |
| 2022/01/30 21:00 | 48 | 66 | Inf. |

Table Notes

[a.] This table is analogous to the first four columns of Table 4, but with UT dates for additional events enabled by doubling the large size scenario. We list these dates because events could conceivably occur on them if the components happen to be highly elongated (or doubled) and oriented in favorable directions on these dates. Also included here are dates omitted from Table 4 due to unfavorably low solar elongations.

## Conclusion

The two components of the transneptunian binary system (385446) Manwë and Thorondor orbit one another with a period of 110.176 ± 0.018 days, a semimajor axis of 6674 ± 41 km, and an eccentricity of 0.5632 ± 0.0070.  The plane of their mutual orbit sweeps across the inner solar system twice during each three century long heliocentric orbit, with the next such passage being anticipated during the next few years.  This special geometry provides opportunities to observe mutual events, when as seen from Earth, the two bodies take turns occulting and/or eclipsing one another.  Mutual events offer a powerful tool to investigate a binary system's physical parameters in much greater detail than can normally be done for such small, distant objects.  Observations of mutual events can constrain the sizes and thus bulk densities of the bodies, along with their shapes, and even potentially identify albedo patterns on their otherwise unresolvable surfaces.  However, mutual event studies of the Manwë and Thorondor system present a number of challenges, as this paper describes.  First, there are relatively few observable mutual events, owing to the small sizes of the objects relative to their separation, along with the long period of their mutual orbit.  Second, the events have long durations, necessitating use of a space-based observatory and/or coordination between multiple ground-based telescopes to observe any event in its entirety.  Third, the two bodies are unlikely to have their rotation states tidally locked to their orbital period, and their apparently large amplitude lightcurves suggest that the profile each presents could be highly variable, complicating interpretation of event lightcurves.  The best hope may be in an observing campaign in which telescopes at multiple longitudes coordinate to observe as many events as possible.  With an ensemble of event lightcurves, along with out-of-event photometric monitoring, it could be possible to simultaneously refine the mutual orbit parameters, while solving for the objects' sizes, shapes, spin states, and center-to-limb photometric behaviors.

## Acknowledgments

This work is based on NASA/ESA Hubble Space Telescope program 13404.  Support for this program was provided by NASA through a grant from the Space Telescope Science Institute (STScI), operated by the Association of Universities for Research in Astronomy, Inc., under NASA contract NAS 5-26555.  We are especially grateful to Linda Dressel and Tony Roman at STScI for their help in designing and scheduling the observations.  This manuscript benefited from insightful and constructive reviews by D. Ragozzine and J. Berthier whom we thank for their efforts. We also thank the creators and maintainers of the NASA/JPL NAIF/SPICE ephemeris system (http://naif.jpl.nasa.gov/naif/), the Johnston archive (http://www.johnstonsarchive.net/astro/asteroidmoons.html), and Lowell Observatory's astorb.dat database and associated infrastructure (ftp://ftp.lowell.edu/pub/elgb/ astorb.html). Finally, we thank the free and open source software communities for empowering us with key tools used to complete this project, notably Linux, the GNU tools, LibreOffice, Python, MariaDB, Evolution, and FVWM.